\documentclass[12pt]{article}
\usepackage{graphicx}
\usepackage{epsfig}

\newcommand\pubnumber{Article 24 in eConf C1304143}
\newcommand\pubdate{\today}

\def\cofc{Dept. Physics and Astronomy, The College of Charleston, \\ Charleston, SC  29424-0001}
\def\cornell{Dept. Astronomy, Cornell University, Ithaca, NY 14853}
\def\duke{Dept. Statistical Science, Duke University, Durham, NC  27708}
\def\uah{Dept. Physics, University of Alabama in Huntsville, \\ Huntsville, AL 35899}
\def\support{\footnote{This work is supported by NASA ROSES AISR program grant NNX09AK60G. R. Preece acknowledges support from GBM through NNM11AA01A/MSFC.}}

\def\Title#1{\begin{center} {\Large #1 } \end{center}}
\def\Author#1{\begin{center}{ \sc #1} \end{center}}
\def\Address#1{\begin{center}{ \it #1} \end{center}}

\newcommand\pubblock{\rightline{\begin{tabular}{l} \pubnumber\\
         \pubdate  \end{tabular}}}
\newenvironment{Abstract}{\begin{quotation}  }{\end{quotation}}
\newenvironment{Presented}{\begin{quotation} \begin{center} 
             PRESENTED AT\end{center}\bigskip 
      \begin{center}\begin{large}}{\end{large}\end{center} \end{quotation}}

\def\beq{\begin{equation}}
\def\eeq#1{\label{#1}\end{equation}}
\def\eeqn{\end{equation}}

\def\beqa{\begin{eqnarray}}
\def\eeqa#1{\label{#1}\end{eqnarray}}
\def\eeqan{\end{eqnarray}}

\let\bar=\overbar

\def\Dslash{\not{\hbox{\kern-4pt $D$}}}
\def\dslash{\not{\hbox{\kern-2pt $\del$}}}

\def\msb{{\bar{\ssstyle M \kern -1pt S}}}

\begin{document}
\begin{titlepage}
\pubblock

\vfill
\Title{A template for describing intrinsic GRB pulse shapes}
\vfill
\Author{ Jon Hakkila\support}
\Address{\cofc}
\Author{ Thomas J. Loredo$^1$}
\Address{\cornell}
\Author{ Robert L. Wolpert$^1$}
\Address{\duke}
\Author{ Mary E. Broadbent$^1$}
\Address{\duke}
\Author{ Robert D. Preece$^2$}
\Address{\uah}
\vfill
\begin{Abstract}
A preliminary study of a set of well-isolated pulses in GRB light curves indicates that simple pulse models, with smooth and monotonic pulse rise and decay regions, are inadequate.  Examining the residuals of fits of pulses to such models suggests the following patterns of departure from the smooth pulse model of Norris et al. (2005):  A {\em Precursor Shelf} occurs prior to or concurrent with the exponential {\em Rapid Rise}. The pulse reaches maximum intensity at the {\em Peak Plateau}, then undergoes a {\em Rapid Decay}. The decay changes into an {\em Extended Tail}. Pulses are almost universally characterized by hard-to-soft evolution, arguing that the new pulse features reflect a single physical phenomenon, rather than artifacts of pulse overlap.
\end{Abstract}
\vfill
\begin{Presented}
``Huntsville in Nashville'' Gamma Ray Burst Symposium\\
Nashville, Tennessee,  April 14--18, 2013
\end{Presented}
\vfill
\end{titlepage}
\def\thefootnote{\fnsymbol{footnote}}
\setcounter{footnote}{0}

\section{Introduction}

Gamma-ray burst (GRB) pulses exhibit highly correlated temporal and spectral properties. A review of pulse properties may be found in Hakkila et al.~2013, in preparation, but are summarized briefly here. Pulse peak flux, duration, fluence, hardness, asymmetry, and lag are highly correlated/anti-correlated.  Pulse lag inversely correlates with pulse peak luminosity, which is the basis of the lag vs. luminosity relation in integrated prompt emission.  Pulses exhibit longer decay than rise rates (temporal asymmetry), hard-to-soft spectral evolution, and longer durations at lower energies than at higher ones. 

Pulse property correlations are remarkable, given that they can be easily observed in photon counts in the observer's frame. The large distances of GRBs (redshifts of $z \ge 1$) suggest that any intrinsic correlations should be smeared out or distorted by effects of observational cosmology (e.g.~time dilation, the inverse square law, K-corrections) and instrumental effects. Instead, the processes by which and environments in which pulses originate (e.g.~from relativistic ejected material traveling at Lorentz factors of $\approx 300$) seem to overpower the effects resulting from universal expansion and instrumental biases.

Correlated pulse properties have been observed in many GRB classes and in a range of environments. They have been measured in the Long and Short burst classes in BATSE bursts, in the Short class of Swift bursts, in GRB pulses observed by HETE-2 and GBM, in x-ray flares observed in Swift afterglows, and in optical flares.

%Correlated pulse properties have been observed in a variety of GRB types and in a range of environments. They have been measured in the Long and Short burst classes in BATSE bursts (e.g. \cite{hak11}), in the Short class of Swift bursts \citep{nor11}, in GRB pulses observed by HETE-2 \citep{ari10}, in x-ray flares observed in Swift afterglows (e.g. \cite{chi10, mar10}), and in optical flares (e.g. \cite{li12}).

The similarity among GRB pulses seems to be simple yet constraining in terms of physical models. How similar or different are the shapes of GRB pulses? 
%This work in progress addresses these issues.

\section{Methodology}

To address this question we have assembled a database consisting of 48 pulses from 47 GRBs in the 20 keV to 1 MeV energy range. Most of these are BATSE pulses being used to compile a GRB pulse catalog, and three of the pulses are from GBM. The three GBM pulses have also been fitted in Suzaku data, but we choose to include only the higher signal-to-noise GBM fits in this analysis.

We register light curves with durations and amplitudes spanning orders of magnitude by fitting each light 64-ms curve to the 4-parameter pulse model of \cite{nor05}: 

\begin{equation}
I(t) = A \lambda \exp^{[-\tau_1/(t - t_s) - (t - t_s)/\tau_2]},
\end{equation}
where $t$ is time since trigger, $A$ is the pulse amplitude, $t_s$ is the pulse start time, 
$\tau_1$ and $\tau_2$ are characteristics of the pulse rise and pulse decay, and the constant 
$\lambda = \exp{[2 (\tau_1/\tau_2)^{1/2}]}$. The pulse peak time occurs at time $\tau_{\rm peak} = t_s + \sqrt{\tau_1 \tau_2}$. 

We fit the pulse data to this model with the iterative procedure described in \cite{hak11}.

Observable pulse parameters can be obtained from the four free pulse fit parameters. Measures of pulse duration $w_n$ can be defined in terms of the times when the fitted intensity has dropped to $e^{-n}$ of its maximum value, and are given by
\begin{equation}
w_n = \tau_{\rm rise} + \tau_{\rm decay} =  n \tau_2 \sqrt{1+4\mu/n}.
\end{equation}
The asymmetry $\kappa_n$ is 
\begin{equation}
\kappa_n = 1/\sqrt{1+4\mu/n)}.
\end{equation}
We define the {\em base pulse duration} $w$ and the {\em base asymmetry} $\kappa$ by the time interval between instances at which the intensity is $e^{-3}$ of its maximum value ($4.98\%$), or 
\begin{equation}
w = \tau_2 [9 + 12\mu]^{1/2};    \kappa = [1 + 4\mu/3]^{-1/2}.
\end{equation} 

The asymmetry ranges from $\kappa=0$ for a symmetric pulse to $\kappa=1$ for an asymmetric pulse with a rapid rise and slow decay (this is a correction to the asymmetry definitions given in our previous pulse-fitting papers (e.~g.~\cite{hak08, hak09,hak11}): the published asymmetry values $\kappa_{\rm prev}$ can be approximately transformed to the new value with $\kappa \approx 0.06 e^{2.8 \kappa_{\rm prev}}$.

In fitting GRB pulses from BATSE, GBM, Suzaku, and Swift, we have discovered that the Norris et al.~model systematically underestimates flux at and near the pulse peak, while overestimating it immediately following the pulse peak. We decide to combine the residuals of several pulse fits to increase our signal-to-noise, and to potentially determine whether this can be used to make systematic improvements to the Norris et al.~pulse shape. However, since GRB pulses have durations and peak fluxes spanning many orders of magnitude (and since they are observed with a variety of instruments having different signal-to-noise responses), we have rescaled the residuals to a normalized pulse duration and to the largest residual found within the duration range before summing and averaging them.

\section{Analysis}

The mean residuals of the 48 pulses are shown in Figure 1a, plotted on a fiducial timescale corresponding to the base duration. The solid curve shows the mean residual; the sum of this and the Norris shape is essentially a global template pulse shape.  The vertical error bars (bounded by tic marks) indicate how well the template is determined (1 standard deviation).Ê The vertical lines (no tics) display the RMS range of departure from the mean shape across the population of pulses. The residuals display trends deviating significantly from zero, supporting our hypothesis that the true pulse shape differs systematically from the Norris et al.~pulse model. However, the deviations from the model are small, indicating that the Norris et al.~pulse shape is a good first-order approximation to the true pulse shape. The pulse model indeed underestimates the true pulse shape near the pulse peak and overestimates it immediately following it, but also exhibits additional flux at the beginning of the pulse rise and on the pulse decay. The fitted pulse peak (indicated by the vertical dashed line) is found at a normalized phase of $0.191.$

To better assess the extent to which the true pulse shape differs from the model, we redefine the pulse duration using a decay time of $w_{8}$, then extend this time earlier than the formal pulse start time $t_s$ by an amount equal to one-tenth the pulse decay time. In other words, since
\vspace{-1 mm}
\begin{equation}
w_{8} = 8 \tau_2 \sqrt{1 + \mu/2},
\end{equation}

the new fiducial timescale $w_{\rm fiducial}$ is defined to be

\begin{equation}
w_{\rm fid} = 4.4 \tau_2 [\sqrt{1+\mu/2}+1] + \sqrt{\tau_1 \tau_2}.
\end{equation}

Figure 1b shows the residuals for the larger fiducial timescale defined in Equation 6. On this timescale, the mean pulse start time occurs at phase $0.110$, while the mean fitted pulse peak is found at phase $0.237$. There are three excesses in the mean residuals, found during distinct pulse phases. The first of these is prior to the pulse rise, the second is near the pulse peak, and the third on the pulse decay. Between these are two well-defined dips, where the flux drops well below the fitted pulse. When combined, the residual excesses and depletions give the appearance of a wave.

In general, these corrections to GRB pulse curves indicate that a typical pulse starts before the Norris et al.\ pulse model predicts it should, and initially brightens more slowly than the model predicts (during the {\em Precursor Shelf} phase). The intensity subsequently increases rapidly (during the {\em Rapid Rise} phase, but slows before reaching a brighter {\em Peak Plateau}, which occurs slightly later than the model predicts. The intensity then decreases rapidly in the {\em Rapid Decay} phase, but undergoes a re-brightening during the {\em Extended Tail} phase; this maintains its brightness for a significantly long time. The net effect demonstrated by the residuals is that a single pulse light curve has several distinct variations, rather than following a single smooth temporal evolution. We also note that there is diversity in the behavior in the precursor shelf segment---sometimes this is faint, while at other times it is quite bright; "shelf" describes the ensemble-averaged behavior only.

%%%%%%%%%%%%%%%%%%%%%%%%%%%%%%%%%%%%%%%%%%%%%%%%%%%%%%%%%%%%%%%%%%%%%%%%%
%%
%%   use this format to include an .eps figure into your paper
%%

%\begin{figure}[htb]
%\centering
%\includegraphics[height=2.0in]{resids_3.eps}
%\caption{Scaled residuals for the sample using the standard duration defined by $w_3$.}
%\label{fig:f1}
%\end{figure}

\begin{figure}
\centering
\begin{minipage}{.5\textwidth}
  \centering
  \includegraphics[width=3in]{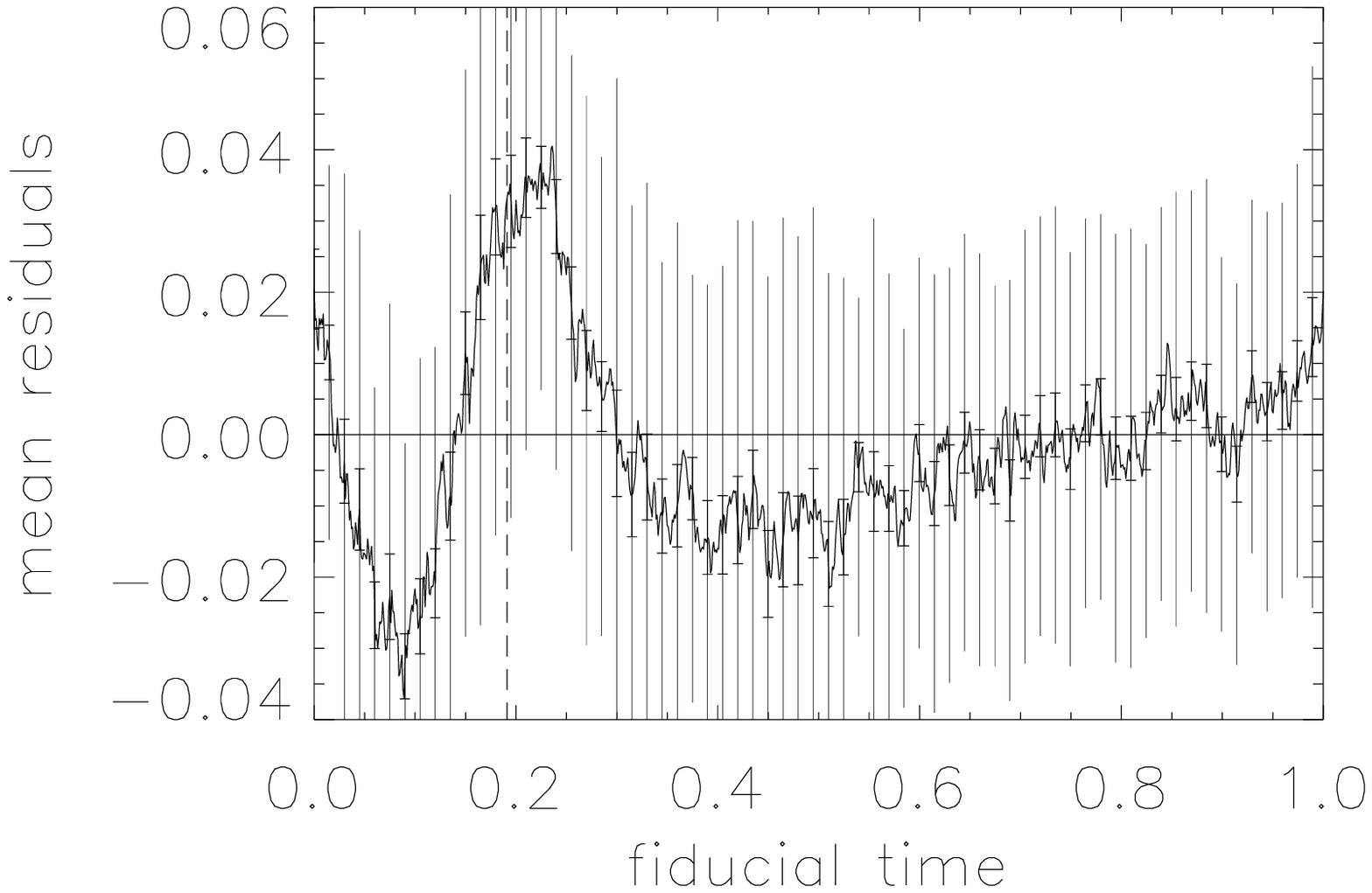}
%  \captionof{Figure 1a.Scaled residuals for the sample using the standard duration defined by $w_3$.}
%  \label{fig:f1}
\end{minipage}%
\begin{minipage}{.5\textwidth}
  \centering
  \includegraphics[width=3in]{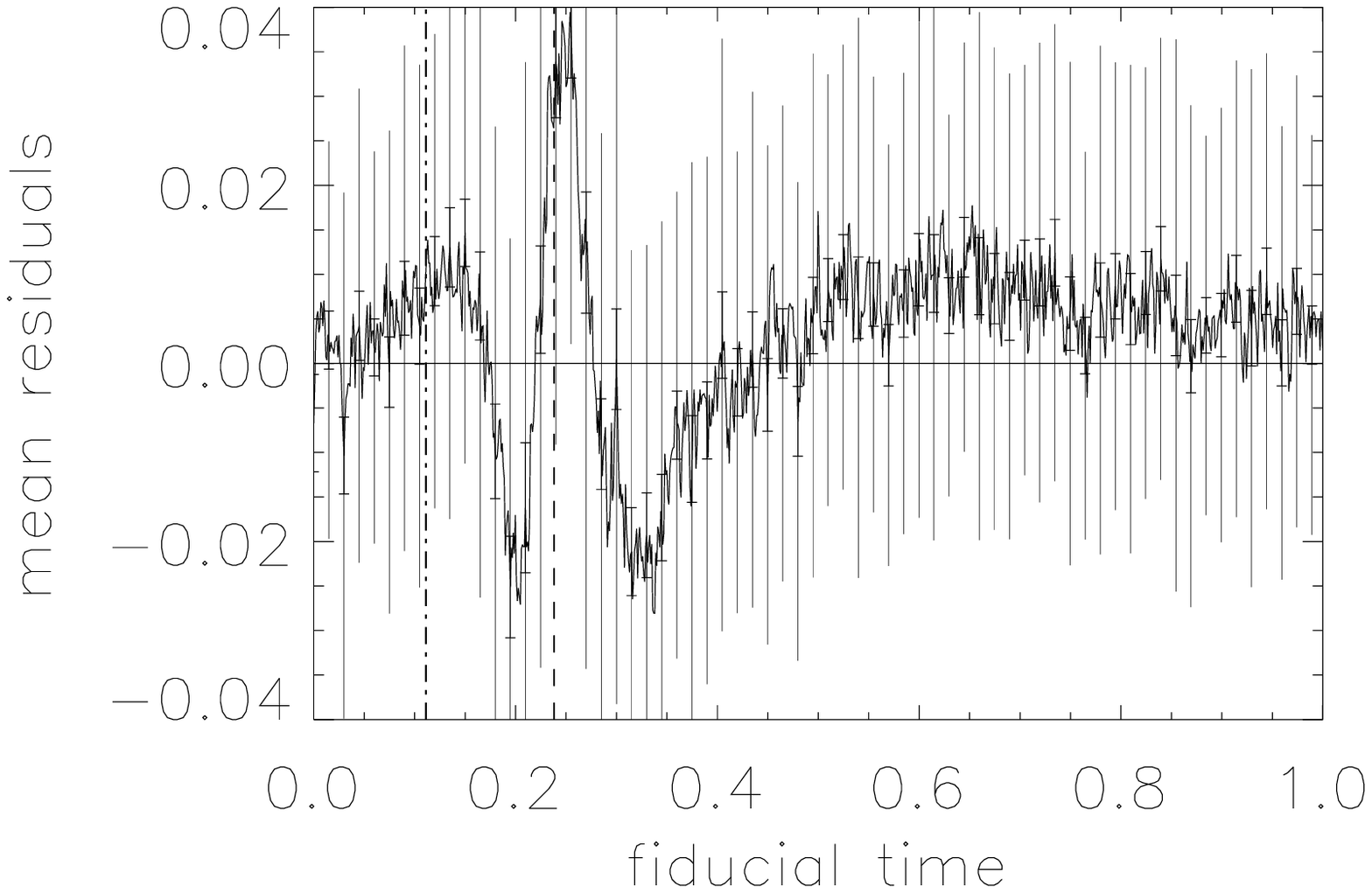}
%  \captionof{Figure 1b. Scaled residuals for the sample using the standard duration defined defined in Equation 6.}
%  \label{fig:test2}
\end{minipage}
\caption{Scaled residuals for the sample using the standard duration $w_3$ (left panel) and the duration defined in Equation 6 (right panel).}
\label{fig:f1}
\end{figure}

\begin{figure}
\centering
\begin{minipage}{.5\textwidth}
  \centering
  \includegraphics[width=3in]{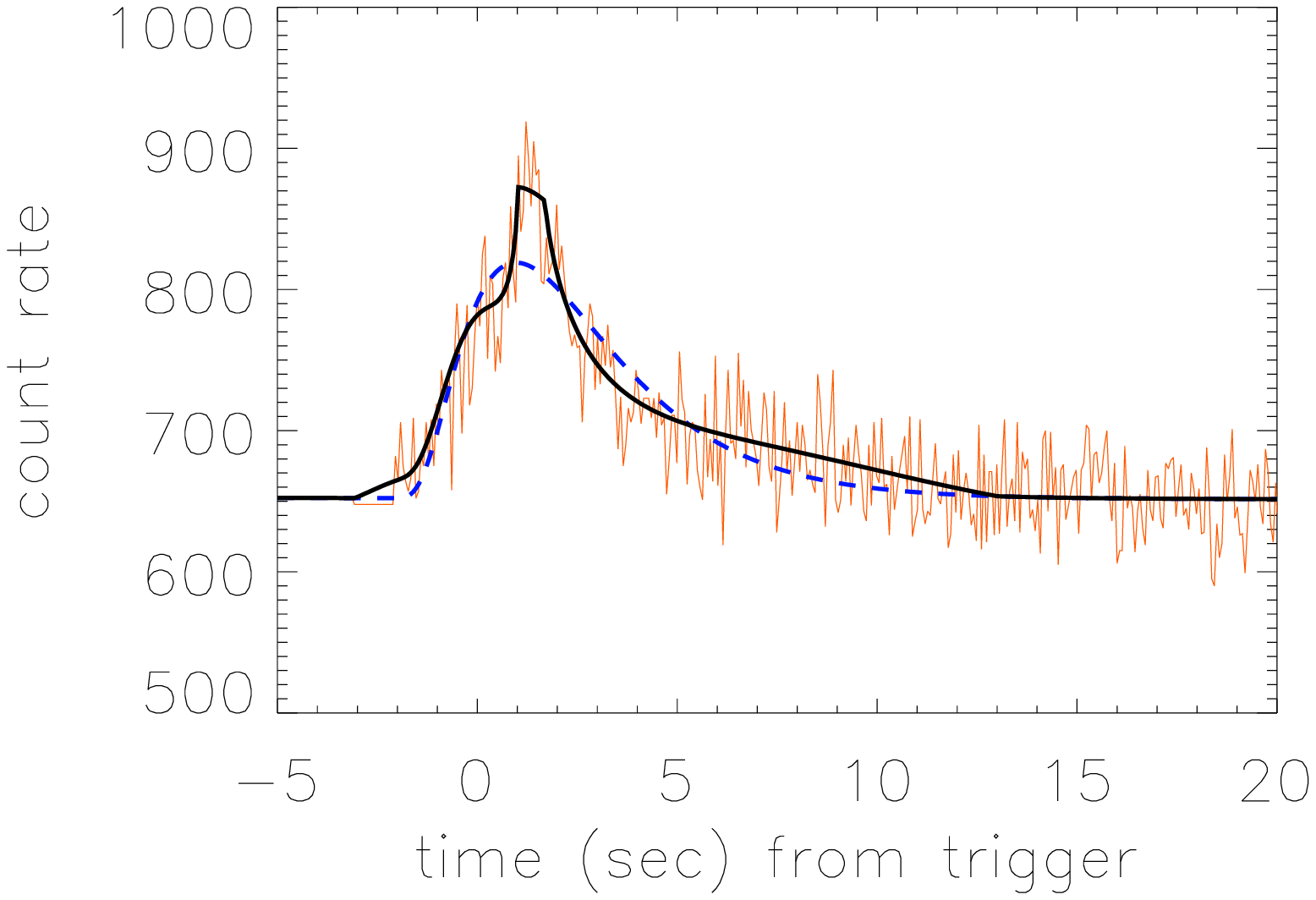}
%  \captionof{Figure 1a.Scaled residuals for the sample using the standard duration defined by $w_3$.}
%  \label{fig:f1}
\end{minipage}%
\begin{minipage}{.5\textwidth}
  \centering
  \includegraphics[width=3in]{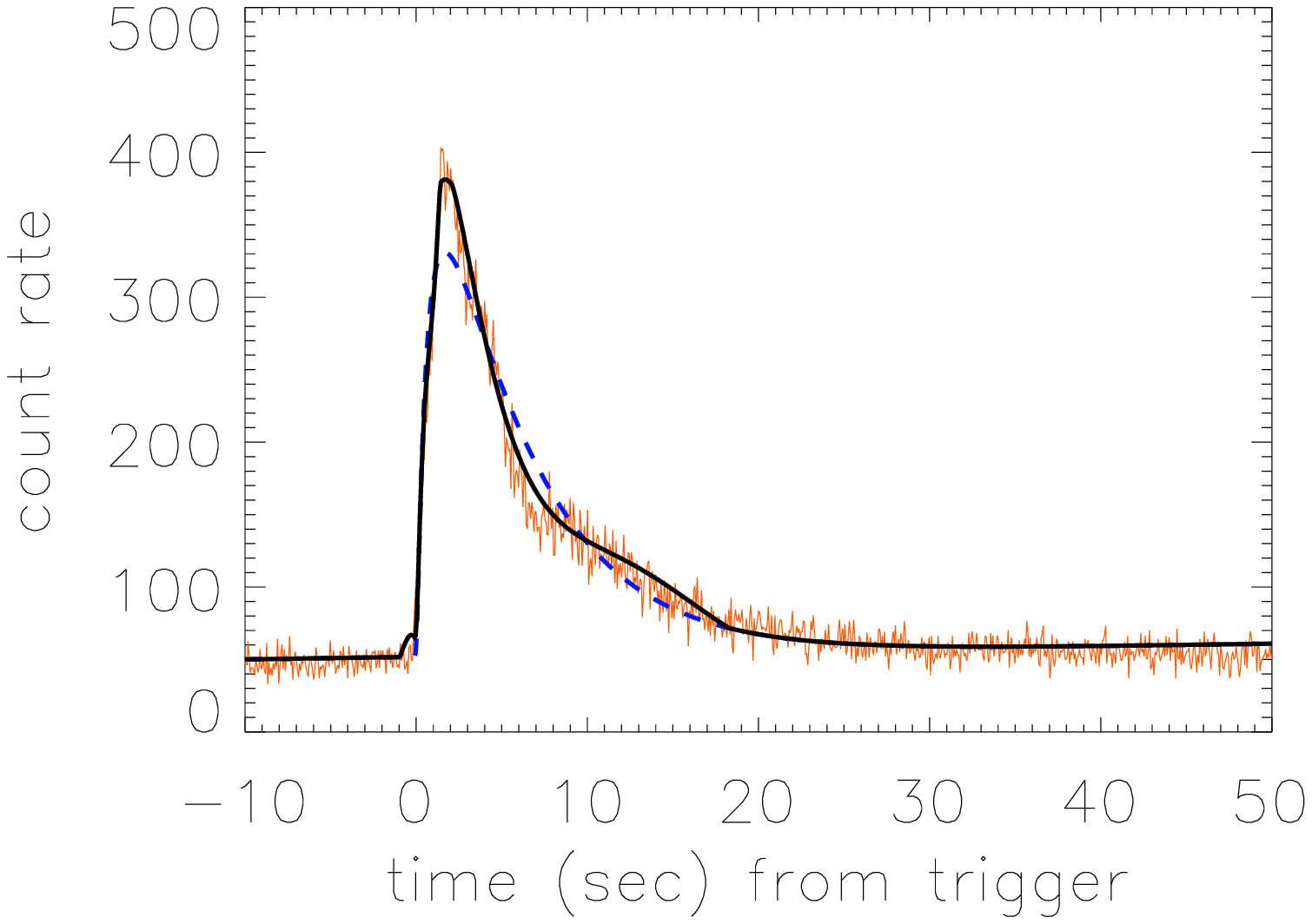}
%  \captionof{Figure 1b. Scaled residuals for the sample using the standard duration defined defined in Equation 6.}
%  \label{fig:test2}
\end{minipage}
\caption{Fitted pulses in BATSE trigger 3026 (left panel) and GRB 100707a (right panel). The Norris et al.~fit (dashed line) is augmented by the template (solid line).}
\label{fig:f2}
\end{figure}

\begin{figure}
\centering
\begin{minipage}{.5\textwidth}
  \centering
  \includegraphics[width=3in]{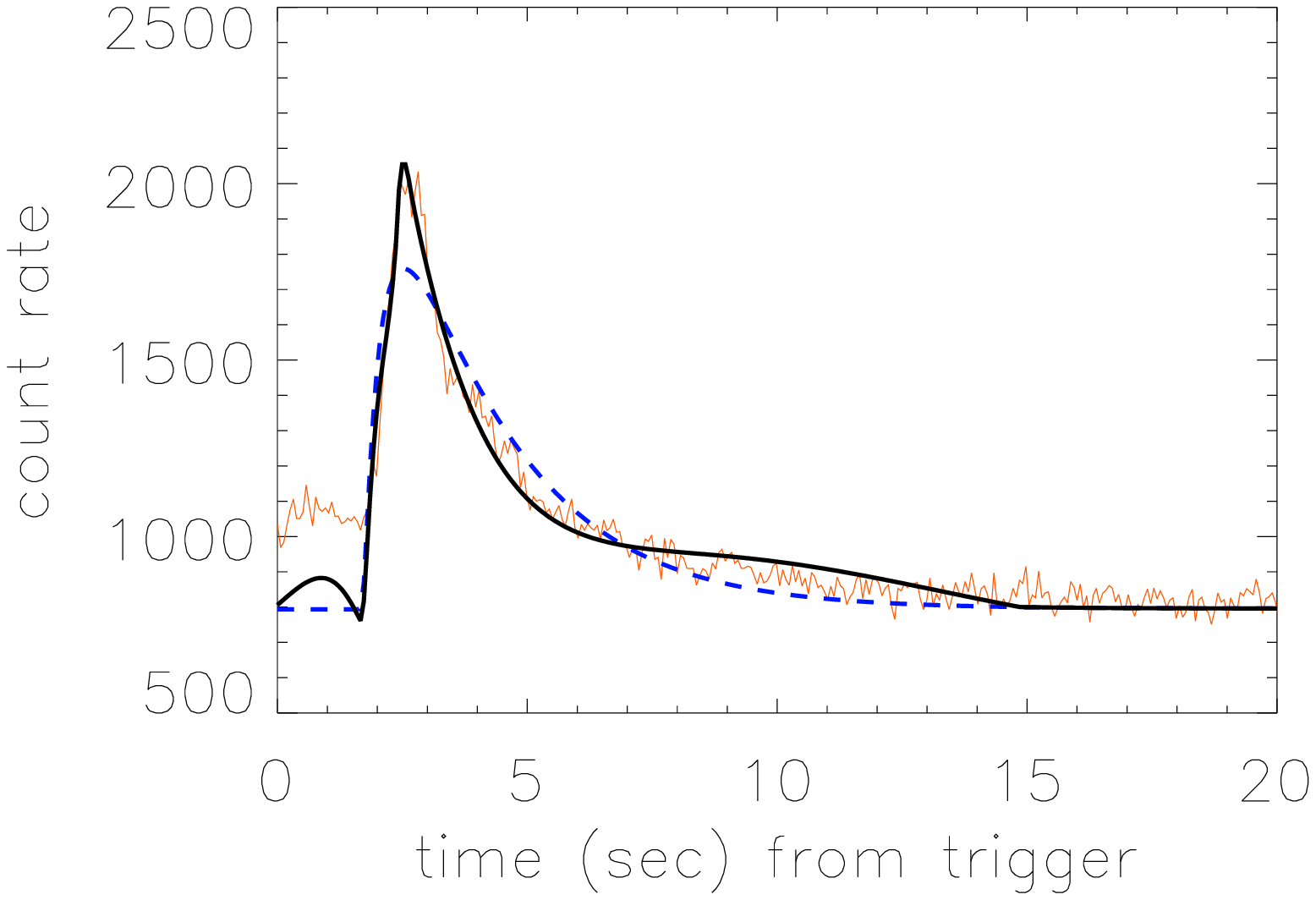}
%  \captionof{Figure 1a.Scaled residuals for the sample using the standard duration defined by $w_3$.}
%  \label{fig:f1}
\end{minipage}%
\begin{minipage}{.5\textwidth}
  \centering
  \includegraphics[width=3in]{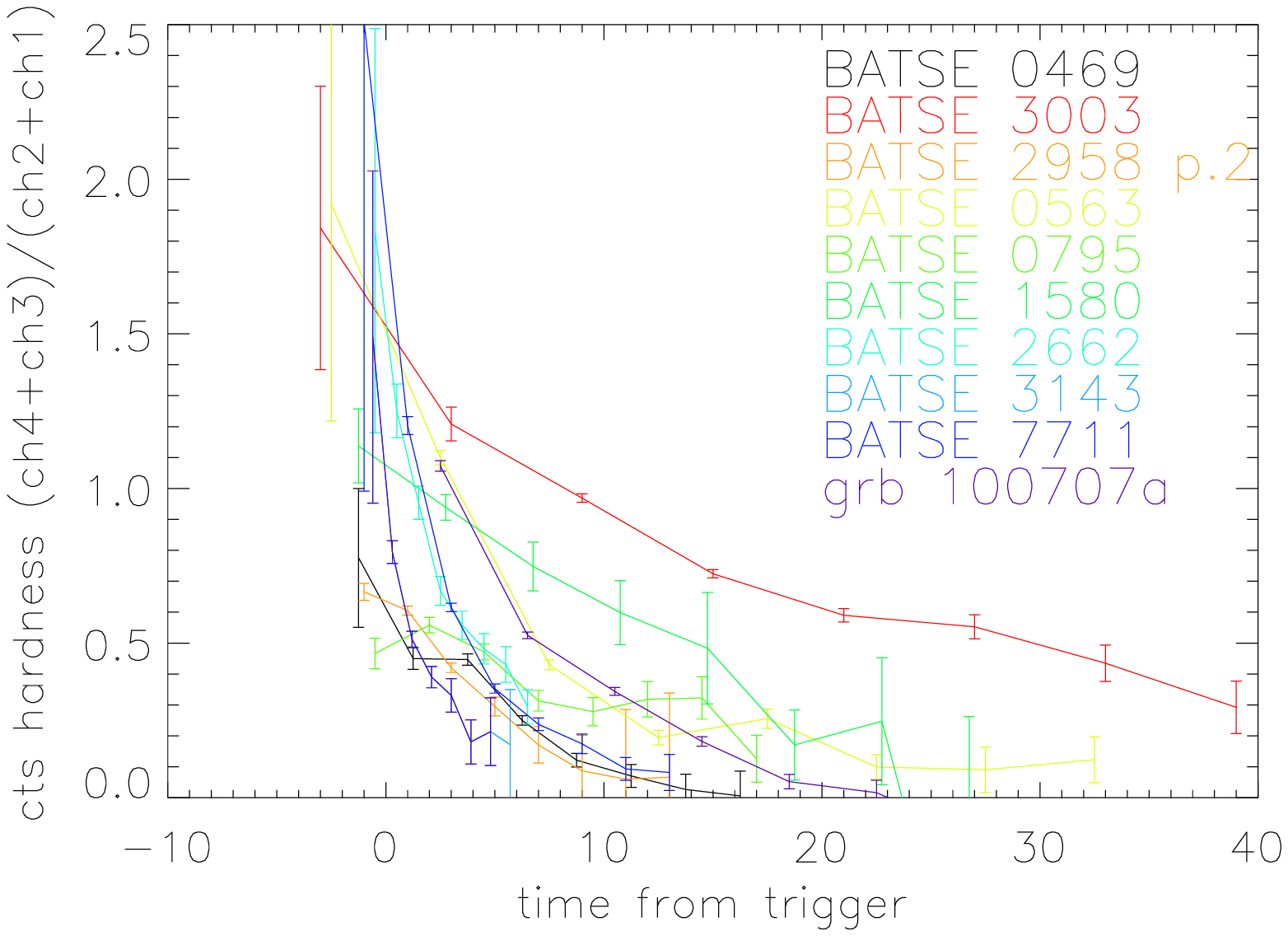}
%  \captionof{Figure 1b. Scaled residuals for the sample using the standard duration defined defined in Equation 6.}
%  \label{fig:test2}
\end{minipage}
\caption{Fitted pulse in BATSE trigger 0469 (left panel), and counts hardness for a GRB pulse sample (right panel).}
\label{fig:f2}
\end{figure}

%\begin{figure}[htb]
%\centering
%\includegraphics[height=2.0in]{BATSE0469.eps}
%\caption{Fitted pulse in BATSE trigger 0469. The phases of the residual excesses and depletions in this pulse support the idea that it is a single pulse with a bright precursor shelf, rather than several overlapping pulses}
%\label{fig:f3}
%\end{figure}

%%%%%%%%%%%%%%%%%%%%%%%%%%%%%%%%%%%%%%%%%%%%%%%%%%%%%%%%%%%%%%%%%%%%%%%%%%%

In theory, we should be able to combine the template with the Norris et al.~pulse fits to reproduce the GRB pulse light curves in our sample. However, when we try to make corrections to individual pulse light curves, we find that the mean residuals do not always align at the same time during the fiducial scale with the individual pulse residuals.  By inspection, we notice that the pulse residuals depend on pulse asymmetry, and subdivide our GRB sample into four separate, asymmetry-dependent subsamples in an attempt to develop more appropriate templates. The subsamples are composed of (a) 12 symmetric pulses having $\kappa < 0.45$ ($\langle \kappa \rangle = 0.27$), (b) 10 slightly asymmetric pulses having $0.45 \le \kappa < 0.67$ ($\langle \kappa \rangle = 0.58$), (b) 13 asymmetric pulses having $0.67 \le \kappa < 0.81$ ($\langle \kappa \rangle = 0.73$), and (c) 12 very asymmetric pulses having $\kappa \ge 0.81$ ($\langle \kappa \rangle = 0.87$).  A smoothed template is produced from each subsample.

With these asymmetry-dependent templates, we have improved the fits of several GRB pulses. The results are shown in Figures 2a and 2b for BATSE trigger 3026 ($\kappa=0.56$) and GRB 100707a ($\kappa=0.83$).  The results show that, when combined with the Norris et al.~pulse model, the asymmetry-dependent templates allow us to more accurately describe the pulse data than can be done using the mean pulse template. 

The results also demonstrate that the phase of the first residual wave peak shifts as pulse asymmetry changes: symmetric pulses exhibit this peak prior to the fitted pulse start time $t_s$ while asymmetric pulses exhibit the peak later, on the pulse rise.  The fluctuations can sometimes combine to be seen as one to three separate peaks within the main pulse. This goes against the notion that a pulse should be defined by a single pulse peak and a smoothly-evolving light curve. To demonstrate the efficacy of this statement, we provide a fit and correction to BATSE trigger 469 in Figure 3a. Prior to this analysis, this pulse was most appropriately fitted by three separate pulses rather than by one single one.

Finally, we supply supporting evidence that the precursor shelf is indeed an important part of each GRB pulse. We plot a counts hardness for ten GRB pulses in Figure 3b; this is the ratio of the 100 keV - 1MeV counts to the 20 keV - 100 keV counts. The first bin in each plot contains the hardness of the precursor shelf, the second and third bins typically contain hardness measurements from the pulse rise, and the remainder of the bins contain hardness during the pulse decay. Pulses generally exhibit hard-to-soft evolution, with the precursor shelf containing the highest energy photons emitted by the pulse.

\section{Conclusions}

This work in progress has uncovered a template-like component in the light curves of isolated GRB pulses:Ê pulses with durations and amplitudes spanning orders of magnitude appear to have very similar shapes after a simple pre- and post-peak temporal rescaling. The characteristics of the template are asymmetry-dependent.

This finding has important implications for data-driven GRB pulse analysis. ÊInterpretation of pulse decomposition results depends critically on adopting pulse shapes that accurately reflect the spectro-temporal behavior of a real physical pulse. Such findings likely also have important implications for phenomenological and physical modeling of GRBs.  The specific features we find potentially constrain such models in several ways:

\begin{itemize}
\item Physical models must account for the approximate uniformity of pulse shapes.
\item Although the flux rises and falls dramatically over the course of a pulse, the hardness decreases nearly monotonically; physical models must account for this surprising spectro-temporal behavior.
\item Physical models must explain the precursor shelf.
\item Physical models must explain why the extended pulse tail produces faint emission as long as 200-300 seconds after the main pulse (this might be the extended GRB emission seen by \cite{con02}).
\end{itemize}

%This exploratory analysis in the spirit of {\em functional data analysis} (FDA) methods developed in statistics, and our work points to the potential value of FDA for GRB pulse studies. 
These results and conclusions are guiding our own ongoing pulse decomposition work. We thank Demos Kazanas for helpful discussions.

\end{document}